\documentclass[conference]{IEEEtran}

\usepackage[pdftex]{graphicx}
\usepackage[usenames,dvipsnames,table,xcdraw]{xcolor}
\usepackage{array}
\usepackage{multirow}
\usepackage[normalem]{ulem}
\useunder{\uline}{\ul}{}
\usepackage{booktabs}
\usepackage{supertabular,booktabs}
\usepackage{longtable}
\usepackage{lscape}
\usepackage{cite}
\usepackage{url}
\usepackage{hyperref}
\usepackage{dirtytalk}
\usepackage{graphicx}
\usepackage{comment}
\usepackage{makecell}
\usepackage{svg}
\usepackage{amsmath}
\usepackage[T1]{fontenc}

\usepackage[caption=false]{subfig}
\graphicspath{{Images/}}
\usepackage[normalem]{ulem}
\useunder{\uline}{\ul}{}

\usepackage{tabularx}
\usepackage{balance}
\usepackage{nameref}

\usepackage[most]{tcolorbox}

\usepackage{xcolor}          

\definecolor{lightblue}{RGB}{0, 0, 100}

\newtcolorbox{MyBox}{
  colback=white,
  colframe=lightblue,
  fonttitle=\bfseries,
  coltitle=black,
  sharp corners,
  boxrule=1pt,
  left=5pt,
  right=5pt,
  top=5pt,
  bottom=5pt,
  breakable
}

\definecolor{purplish}{HTML}{D8DFE3}
\definecolor{purplishlight}{HTML}{EBEFF3}
\definecolor{purplishdark}{HTML}{FF7F50}

\newtcolorbox[auto counter,number within=section]{rqbox}[2]{
    nameref=#1,
    title=\small{#1}, 
    enhanced,
    attach boxed title to top left={yshift=-6pt, xshift=8pt},
    boxed title style={size=small,boxsep=1pt},
    colframe=purplishdark,colback=white,colbacktitle=purplishdark,
    boxsep=2pt,left=2pt,right=2pt,top=6pt,bottom=2pt,middle=2pt
}

\newcommand{\rqtextone}{How are LLM-powered systems tested, and what challenges arise during their verification and validation processes?}

\begin{document}

\title{Testing the Untestable? An Empirical Study on the Testing Process of LLM-Powered Software Systems}


\author{
\IEEEauthorblockN{Cleyton Magalhaes}
\IEEEauthorblockA{UFRPE\\
 Recife, PE, Brazil \\
cleyton.vanut@ufrpe.br}
\and

\IEEEauthorblockN{Italo Santos}
\IEEEauthorblockA{University of Hawai‘i at Mānoa\\
 Honolulu, Hawai‘i, USA \\
 italo.santos@hawaii.edu}
\and

\IEEEauthorblockN{Brody Stuart-Verner}
\IEEEauthorblockA{University of Calgary\\
Calgary, AB, Canada \\
brody.stuartverner@ucalgary.ca} 
\and

\IEEEauthorblockN{Ronnie de Souza Santos}
\IEEEauthorblockA{University of Calgary\\
Calgary, AB, Canada \\
ronnie.desouzasantos@ucalgary.ca} 

}


\IEEEtitleabstractindextext{%
\begin{abstract}
\textit{Background:} Software systems powered by large language models are becoming a routine part of everyday technologies, supporting applications across a wide range of domains. In software engineering, many studies have focused on how LLMs support tasks such as code generation, debugging, and documentation. However, there has been limited focus on how full systems that integrate LLMs are tested during development. \textit{Aims:} This study explores how LLM-powered systems are tested in the context of real-world application development. \textit{Method:} We conducted an exploratory case study using 99 individual reports written by students who built and deployed LLM-powered applications as part of a university course. Each report was independently analyzed using thematic analysis, supported by a structured coding process. \textit{Results:} Testing strategies combined manual and automated methods to evaluate both system logic and model behavior. Common practices included exploratory testing, unit testing, and prompt iteration. Reported challenges included integration failures, unpredictable outputs, prompt sensitivity, hallucinations, and uncertainty about correctness. \textit{Conclusions:} Testing LLM-powered systems required adaptations to traditional verification methods, blending source-level reasoning with behavior-aware evaluations. These findings provide evidence on the practical context of testing generative components in software systems.
\end{abstract}

}

\maketitle

\IEEEdisplaynontitleabstractindextext

\IEEEpeerreviewmaketitle

\section{Introduction}
\label{sec:introduction}

Large Language Models (LLMs) are AI systems capable of understanding and generating human language~\cite{zheng2025towards}. Although their primary purpose lies in natural language processing, LLMs have shown promising performance in several contexts~\cite{rasnayaka2024empirical, hou2024large}. Today, in many domains, software applications are being designed specifically to integrate with LLMs. These integrations enable applications to perform tasks such as querying, content generation, summarization, classification, and decision support~\cite{chen2025empirical, morales2024dsl, feng2024llmeffichecker}. However, building such systems introduces a range of technical challenges, including prompt engineering, model selection, versioning, and unpredictable outputs, that make development and maintenance particularly complex~\cite{chen2025empirical, ma2024my, asgari2024testing}.

Within software engineering, LLMs are being utilized to support a wide range of activities throughout the software development lifecycle. Developers rely on LLMs to write and understand code, debug, generate documentation, and review pull requests~\cite{hou2024large, rasnayaka2024empirical}. LLMs are also increasingly used to support testing activities, including the generation of unit tests, creation of test data, reproduction of bugs, and even validation of expected behavior~\cite{wang2024software, santos2024we, yang2024evaluation}. LLMs' use is especially appealing in tasks that require automation, quick prototyping, or natural language explanations.

In this context, the literature offers numerous studies on how LLMs can support software development tasks, including software testing activities such as test generation, fault localization, and automation~\cite{wang2024software, santos2024we, zapkus2024unit}. However, relatively few studies have examined the testing of systems that embed LLMs as part of their core functionality, despite the increasing prevalence of such applications~\cite{ma2024my, chen2025empirical, asgari2024testing}. Verifying and validating LLM-powered systems presents unique challenges due to their non-deterministic behavior, sensitivity to prompts, and potential for silent updates that alter model behavior over time~\cite{ma2024my, chen2025empirical, asgari2024testing, feng2024llmeffichecker}. To address this gap, this study investigates how LLM-powered systems are tested and identifies the key challenges that teams encounter during this process. We frame our investigation through the following research question:


\newcommand{\rqone}[2][]{
     \begin{rqbox}{\textbf{Research Question}}{#2}
         \rqtextone
         #1
     \end{rqbox}
 }

 \rqone{}

To answer our research question, we analyzed student-led projects developed in a third-year software architecture course. These open-ended, team-based projects required students to design, implement, and test functional LLM-powered applications under conditions that reflected realistic development constraints. While the participants were not professionals, the setting provided insight into how emerging practitioners approach verification and validation when working with generative components, blending established engineering practices with adaptive behaviors shaped by model unpredictability. Our paper delivers three main findings:

\begin{enumerate}
    \item Coding and testing were rooted in source code analysis and manipulation.
    \item Teams engaged in hybrid strategies that combined traditional and model-aware verification and validation practices.
    \item The inherent unpredictability and instability of LLM behavior shaped coding and testing.
\end{enumerate}

This paper is organized as follows. Section~\ref{sec:back} refers to the literature on LLM-powered systems and LLM-powered system testing. Section~\ref{sec:method} describes our methodology. Section~\ref{sec:findings} presents our findings, which we discuss in Section~\ref{sec:discussion}. Section~\ref{sec:limitations} reports the threats to validity. Finally, Section~\ref{sec:conclusions} summarizes our contributions and outlines the directions for future work.

\section{Background} 
\label{sec:back}

This section reviews the literature on the development and testing of LLM-powered systems. We present key findings related to engineering practices and validation challenges reported in multiple studies.

\subsection{Developing LLM-Powered Systems}

LLM-powered systems are being adopted in a variety of domains to support tasks such as classification, summarization, content generation, conversational interaction, and decision support~\cite{chen2025empirical, yang2024talk2care, lin2024arxiv, ramjee2025ashabot}. These systems are typically designed to connect with large models via APIs or embedded engines, enabling language-driven functionality across contexts~\cite{chen2025design, ramjee2025ashabot, lin2024arxiv}. Developers building such systems often follow hybrid processes that combine software engineering with prompt engineering and iterative refinement~\cite{chen2025empirical, cai2025demystifying, chen2025design}.

Unlike traditional systems, LLM-based development workflows rely heavily on trial-and-error cycles, with developers modifying prompts, adjusting temperature and context parameters, and evaluating outputs through informal testing~\cite{cai2025demystifying, chen2025empirical}. This lack of determinism in the development process stems from the models’ black-box behavior and unpredictable responses to even minor prompt changes~\cite{lin2024arxiv}. These characteristics challenge developers' ability to design predictable and robust systems, especially when user trust and safety are involved~\cite{chen2025empirical, chen2025design}.

LLM system developers also face concerns, including selecting appropriate models, handling changes introduced by providers, and ensuring system responses align with user needs and domain constraints~\cite{chen2025design, chen2025empirical}. In practice, alignment strategies often involve fallback logic, multi-step interactions, or human-in-the-loop mechanisms that can catch undesired or inaccurate outputs~\cite{yang2024talk2care, chen2025design}.

Despite the growing importance of these systems, their development remains undersupported by formal design practices or mature tooling, leading most software teams to rely on intuition, community-shared heuristics, and lightweight strategies, with limited standardization across tools and workflows~\cite{ramjee2025ashabot, cai2025demystifying}. This fragmented state has highlighted the need for more structured engineering practices that address the unique challenges of integrating LLMs into software systems~\cite{chen2025empirical}.

\subsection{Testing LLM-Powered Systems}

In the context of software testing, LLM-powered systems introduce challenges that differ from those encountered in traditional software testing. LLM-powered systems are inherently non-deterministic, and even identical prompts can produce different outputs, making activities like regression testing unreliable~\cite{ma2024my}. The frequent update of the model by providers can also lead to behavioral drift, which occurs without changes in the application code, further challenging long-term validations~\cite{ma2024my, chen2025empirical}.

In the literature, some techniques have been proposed to test these systems, including black-box testing, output sampling, behavioral coverage metrics, and prompt-based probing~\cite{asgari2024testing, feng2024llmeffichecker}. However, these techniques are often limited in scale or applicability. As a result, developers frequently rely on manual review or user feedback to identify faults or confirm output quality~\cite{chen2025empirical, chen2025design}.

Testing efforts also face additional challenges, including hallucinations, irrelevant responses, and silent failures that often go undetected without extensive human evaluation~\cite{lin2024arxiv, feng2024llmeffichecker}. In addition, critical aspects such as robustness, fairness, and bias are often under-assessed, especially outside of regulated domains~\cite{huang2024bias, bedi2024systematic}.

Previous literature suggests that testing practices for LLM-powered systems remain ad hoc and not sufficiently formalized. Although some frameworks aim to support more robust evaluations, current adoption is limited. Therefore, there is currently a need for more systematic, repeatable, and context-sensitive validation methods that can address the evolving nature and inherent complexity of LLM applications~\cite{feng2024llmeffichecker, ma2024my, asgari2024testing}.

\section{Method}
\label{sec:method}

This study is an exploratory and descriptive case study~\cite{ralph2020empirical, runeson2009guidelines, yin2017case}, focusing on understanding how undergraduate software engineering students approach the testing of LLM-powered systems. Our goal was to investigate the strategies and challenges that arise when real-world software development intersects with LLM behavior, particularly in settings where instructional content and software practices are closely intertwined.

\subsection{Case and Context}

The case centers on a third-year undergraduate course in software development and architecture offered in a software engineering program. The course covered multiple stages of the software engineering lifecycle, including system design, architectural modeling, database integration, testing practices, and full-stack implementation. While students were exposed to all of these stages, this study focuses specifically on their testing activities and reflections concerning systems that integrate LLMs. 

During the course, students formed project teams and were required to develop and deploy a functional application or website that satisfied three core technical constraints: (1) use of a SQL-based database for persistent storage, (2) accessible deployment through a public or shareable endpoint, and (3) integration of a LLM (e.g., ChatGPT, Gemini, Copilot) as a core system feature. The LLM functionality was not optional; it was expected to deliver a core capability within the system. Common use cases included personalized recommendations, domain-specific content summarization, interactive chatbot services, and natural language interfaces for user interaction.

In preparation for the course project, students received one lecture on prompt engineering and supporting academic materials to introduce LLM-related topics such as model variability, hallucinations, and format mismatches. Although the course primarily focuses on software development and architecture, rather than prompt design, this material was provided to help students reason about prompt behavior when integrating LLMs into their systems. No formal instruction on prompt testing was included, but class materials encouraged teams to iteratively experiment with prompts during development.

Software testing, on the other hand, was explicitly addressed in the course through instructional materials and formed part of the evaluation criteria. Students received formal instruction in diverse testing techniques, including unit and system testing, exploratory and scripted approaches, test automation, and behavioral evaluation. These practices were intended to be applied not only to conventional system components (e.g., UI, backend) but also to the outputs produced by the LLMs embedded in their applications.

\subsection{Unit of Analysis}

A total of 21 team projects were completed and deployed by the end of the semester. The teams were composed of 4 to 6 students. As part of their final evaluation, students submitted individual reports reflecting on their team’s testing activities. We collected and analyzed 99 individual reports, which are treated as the unit of analysis in this study. Each report was analyzed independently, as the reflections captured personal testing decisions, observations, and challenges encountered by individual students.

The projects developed spanned a variety of domains, reflecting students’ diverse interests and the open-ended nature of the assignment. Several teams created applications focused on personal health and wellness, including tools for meal planning, calorie estimation, workout guidance, and dietary tracking. Other projects addressed everyday organization and productivity, with features such as financial planning, recipe generation based on available ingredients, and digital wardrobe management. A number of teams developed recommendation systems that supported users in choosing books, movies, computer hardware, or activities tailored to their preferences. Many projects also incorporated elements of image recognition, user profiling, or customization features to improve the user experience. While the specific application areas varied, most systems were interactive, data-driven, and designed for deployment on mobile or web platforms, with attention to usability, responsiveness, and real-time feedback.

LLMs were integrated as core components in all projects, often shaping the central functionality of the system. Teams used LLMs to generate tailored responses, extract information from user input, interpret uploaded data such as images or text, and support conversational or decision-making features. The most commonly used models were Google Gemini and OpenAI’s GPT series. In general, LLMs provided users with advice, suggestions, or recommendations. In others, they generated personalized outputs based on real-time input, such as adapting to user preferences, summarizing content, or explaining results. The LLMs were used both in direct user interactions and as underlying components that informed system logic and behavior.

\subsection{Data Collection}

To ensure research independence and participant protection, the data collection was handled by two teaching assistants who were not involved in the study. After the course ended and final grades were released, these assistants collected all submitted reports and performed the anonymization process. This included the removal of student names, IDs, metadata, and any group identifiers. The reports were then compiled into a single dataset, which was provided to the research team in de-identified form. No sampling was conducted: all 99 anonymized reports were retained and analyzed in full.

\subsection{Data Analysis}

\begin{figure}[t]
\centering
\centerline{\includegraphics[width=0.5\textwidth]{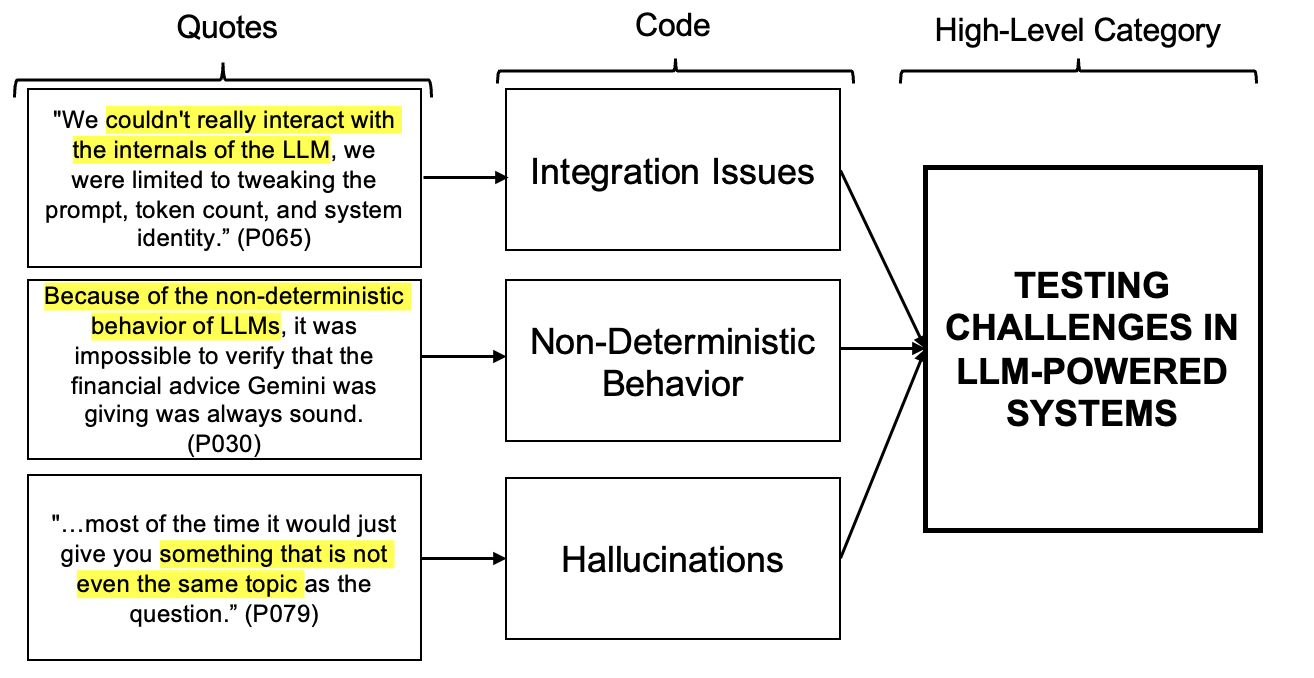}}
\caption{Example of Thematic Analysis Process}
\label{fig:method}
\end{figure}

We conducted a thematic analysis~\cite{cruzes2011recommended} to identify how students described their testing practices in relation to the LLM-powered features. To assist in the coding process, we used ChatGPT 4.0 to support the extraction and classification of narratives across four dimensions related to testing behavior and LLM-specific concerns. Figure~\ref{fig:method} illustrates the thematic analysis process. The following structured prompt was designed to guide the classification:

\begin{MyBox}\small
\textbf{Instruction:} You are assisting in the analysis of student reflections on testing LLM-powered applications. For each report, identify and extract statements that clearly refer to one or more of the following categories: \newline
(1) Type of testing (e.g., exploratory, scripted, automated); \newline
(2) LLM Behavior; \newline
(3) Testing Challenges; \newline
(4) Lessons learned from the testing process. \newline

\textbf{Output Format:} \newline
- Report ID: A unique code (e.g., R015) \newline
- Category: The aspect covered (e.g., “Testing Challenges”) \newline 
- Extracted Snippet: Verbatim text from the report. \newline

\textbf{Note:} Only extract statements where the student makes an explicit claim. Do not infer meaning or assign labels to vague or ambiguous references.
\end{MyBox}

The analysis followed a four-stage process:
\begin{enumerate}
    \item \textbf{Familiarization and prompt calibration}: Researchers reviewed a subset of the reports to refine the identified labels and construct the classification prompt.
    
    \item \textbf{Model-assisted categorization}: ChatGPT-4.0 was used to identify and label relevant excerpts across the 99 reports.
    
    \item \textbf{Manual verification and refinement}: Two researchers independently reviewed all labeled and categorized segments, confirming or adjusting outcomes to ensure contextual accuracy.
    
    \item \textbf{Theme construction}: The categorized excerpts were grouped into thematic patterns that captured how students approached testing LLMs and the obstacles they encountered.
\end{enumerate}

To maintain analytical rigor, we adopted a strict inclusion rule: only statements explicitly made by students were included in the analysis. The researchers did not infer meaning or interpret intent beyond what was clearly expressed in the text. This decision was made to preserve transparency, reproducibility, and consistency in how the data was coded and interpreted.

We selected ChatGPT 4.0 for initial code suggestions to accelerate the open coding process and reduce redundant labeling. Prompts were carefully calibrated by the first author using test excerpts to ensure reliable topic extraction and consistent terminology. All generated codes were reviewed by two researchers independently and validated against the original context. Disagreements were resolved through discussion. We did not rely on the LLM to generate themes or interpret data. Instead, ChatGPT served as a tool to support focused coding, with final coding and thematic synthesis performed manually by the research team. This process followed emerging guidance from recent empirical work discussing the use of LLMs in qualitative analysis, which advocates for transparent roles and human validation when using automated tools to support coding tasks \cite{lecca2024applications}.

Regarding saturation, the dataset offered rich variation across students, and no themes were excluded based on frequency alone. Saturation was monitored iteratively, and by the time 80 reports had been analyzed, no new categories or meaningful sub-themes were emerging. To enhance credibility, two researchers conducted all validation steps collaboratively, resolving any disagreements through discussion. Illustrative quotations are presented throughout the results to support traceability between raw data and emergent findings.

\subsection{Ethical Considerations}
\label{sec:ethics}

This study was conducted in accordance with the institutional ethics guidelines governing research involving human participants. Reports were collected only after the course concluded and final grades were released, ensuring that participation had no impact on student assessment. All data processing and anonymization were performed by teaching assistants who were not involved in the study. The research team had access only to de-identified materials. Due to the personal and course-specific nature of the dataset, the full reports cannot be publicly disclosed. However, anonymized excerpts have been included in this paper to demonstrate the basis for each thematic interpretation and to support the transparency of the results.

\section{Findings} 
\label{sec:findings}

In this section, we present the findings of our analysis, combining thematic coding with excerpts from student reports. The results illustrate how teams approached the task of testing LLM-powered systems, including the strategies they employed to validate behavior, the issues they encountered at both the code and model levels, and the challenges that shaped their testing processes. These findings reveal the layered nature of testing in LLM-integrated applications, offering information on how participants adapted traditional software verification practices to accommodate the unique characteristics of generative models.

\subsection{Validating LLM Behavior: From Code to Output}

The teams applied several testing techniques to verify and validate the behavior of the LLM-powered components in their systems. These techniques addressed different layers of the development process, from testing the code responsible for integrating the LLM into the backend, to manipulating input data and prompt structures, to observing and evaluating the outputs generated by the model. Teams explored how LLM responses behaved under different configurations, ensured that the outputs were handled correctly by the system, and assessed whether those outputs aligned with user expectations and domain requirements. The testing practices varied in structure and depth, depending on the team’s goals and the stage of development. Table~\ref{tab:testing-approaches} includes quotations that illustrate each of the testing techniques applied.

The most commonly reported testing approach was \textit{manual exploratory testing}, where teams interacted with the system to investigate how the LLM responded to various inputs. Unlike ad hoc testing, exploratory testing was conducted with a defined process: teams described following a testing charter or a set of testing goals, often focused on prompting variations, edge cases, or behavioral boundaries. The testing sessions were adaptive but intentional, allowing teams to discover patterns and limitations in LLM behavior while maintaining a structured exploration strategy.

The second most prevalent strategy was \textit{manual ad hoc testing}, in which teams described unplanned and informal testing activities that were conducted without following any structured procedure or predefined objectives. These sessions were largely spontaneous, triggered by new implementation changes or unexpected behaviors, and typically lacked documentation or repeatable steps. Teams relied on intuition and improvisation to test LLM integration, modifying prompts or features in a reactive manner.

A smaller group of teams engaged in \textit{manual scripted testing}, where predefined test cases were used to verify expected system behaviors. These test cases were sometimes documented as checklists or structured scenarios that covered specific functionality. Scripted testing provided a more systematic approach to evaluating whether the LLM output met design or domain requirements, allowing teams to track testing coverage and reproducibility.

Some teams employed \textit{automated unit testing} to validate the smaller, isolated components of the system that interacted with the LLM. These tests were written for backend logic, parsing functions, or other modules involved in preparing or processing LLM data. Unit testing was generally supported by testing frameworks and helped ensure the correctness of low-level functions independent of full-system behavior.

Fewer teams reported using \textit{automated integration testing}, where multiple system components were tested together to confirm that they interacted properly with the LLM. This approach was focused on verifying the correct flow of data between the LLM, database, backend logic, and UI components. Teams used this strategy to ensure that the LLM’s output could be processed, stored, and presented as expected.

While these testing approaches are reported individually for clarity, it is important to note that most teams used a \textit{combination of techniques}. For instance, exploratory testing was frequently used in tandem with unit testing, allowing teams to observe both the system-level responses and the correctness of internal components. Similarly, scripted scenarios were sometimes embedded within larger exploratory sessions, and ad hoc practices emerged when teams encountered unanticipated issues during structured testing. The layering of testing types reflects the complexity of working with LLM-powered systems and the need to validate both predictable and unexpected behavior.

\begin{table}[h]
\caption{Illustrative Quotes by Type of Testing Approach}
\label{tab:testing-approaches}
\centering
\begin{tabular}{p{8cm}}
\toprule
\textbf{Manual: Exploratory} \\
\midrule
\textit{"Our testing technique for the LLM was very exploratory, as we mainly tried different prompts and analyzed the results."} (P001) \\ \\

\textit{"Firstly, early on in the development we used a lot of exploratory testing to try to find the prompt that worked the best for our use case."} (P009) \\ \\

\textit{"Exploratory blackbox testing was done manually and included a test plan for reproducibility."} (P015) \\

\midrule
\textbf{Manual: Ad Hoc} \\
\midrule

\textit{"We would input prompts and manually assess whether the LLM's responses were relevant, accurate, and aligned with the expectations of our application."} (P034) \\\\

\textit{"Also did manual ad hoc testing, like making sure the LLM was giving an output that could be displayed."} (P051) \\ \\

\textit{"When testing LLM responses we mainly testing things manually. During manual testing we had altered our query many times until we received consistent and correct results from the llm."} (P069) \\

\midrule
\textbf{Manual: Scripted} \\
\midrule
\textit{"The primary testing approach we used for our LLM was scripted testing that linked to our requirements traceability matrix."} (P035) \\ \\

\textit{"Functional scripted testing was the primary method used. A test case was written for each different input option and submitted."} (P068) \\ \\

\textit{"We had scripted testing to determine what the test cases were and then from that we had to analyse the returned outputs manually to make sure they followed the prompt."} (P081) \\

\midrule
\textbf{Automated: Unit} \\
\midrule
\textit{"We used unit testing with pytest (python) to validate the LLM's output was consistent and well formatted."} (P021) \\ \\

\textit{"Unit tests: - If user enters all fields, - If user leaves fields empty, - If user spams submit button, etc.."} (P026) \\ \\

\textit{"We used unit testing for the backend because there were certain features that could only be tested using unit testing."} (P041) \\

\midrule
\textbf{Automated: Integration} \\
\midrule
\textit{"To validate the data/outputs coming from the LLM, we primarily experimented with the prompt, and evaluated the output using an integration testing tool like Postman."} (P045) \\ \\

\textit{"Integration tests: Verified the interactions between modules."} (P047) \\ \\

\textit{"Additionally, we performed integration testing to ensure LLM outputs aligned with the current inventory and dietary restrictions."} (P093) \\
\bottomrule
\end{tabular}
\end{table}

\subsection{Issues Identified During LLM Testing: Code-Level and Model Behavior}

The teams reported the main issues and bugs they encountered while verifying and validating the behavior of their LLM-powered systems. These challenges emerged throughout the development process as the students tested the interaction between LLMs and other system components, monitored the results, and iteratively adjusted their code. Table \ref{tab:issue-types} illustrates these findings.

Based on the analysis of student reports, we grouped the identified issues into two distinct categories: \textit{source-level code issues} and \textit{model behavior issues}. This distinction reflected the hybrid nature of LLM-powered systems, where verification involved not only traditional software debugging but also reasoning about the model’s behavior under varying prompt structures and inputs.

Source-level code issues referred to problems rooted in implementation and system design. Several teams encountered \textit{logical errors}, such as incorrect reasoning or flawed backend processing that caused the system to misinterpret otherwise valid LLM responses. \textit{Interface errors} were also common, particularly in the coordination between the front-end, back-end, and the LLM API, which included improperly passed parameters, failures to parse model responses, or breakdowns in how the output was handled and displayed. Although less frequent, some students reported \textit{syntax errors} during LLM integration, particularly when adapting code structures to accommodate API calls or new libraries. A few teams also experienced what we called \textit{teamwork issues}, where miscommunication or poor coordination led to misunderstandings about how prompts should be constructed or how data should flow across the system.

Model behavior issues, on the other hand, emerged from the nature of interacting with a non-deterministic language model. Many students described problems with \textit{LLM output errors}, such as inconsistencies, irrelevant results, or responses that did not align with the prompt intent. These failures often occurred even when the underlying code was functioning as expected, suggesting limitations in the model's reliability. \textit{Prompt sensitivity issues} was another recurring issue: Minor re-writes or formatting changes often produced drastically different outputs, making it difficult to ensure stable and reproducible behavior within the code.

Together, these categories demonstrated the complexity involved in testing LLM-powered systems. Teams were required not only to debug their own code but also to ensure that the integration between system components and LLM behavior was functioning as intended. This often demanded an understanding of the system’s internal logic, data flow, and response handling, while also accounting for the model’s non-deterministic and sometimes unpredictable responses. These factors introduced a layer of behavioral evaluation that extended beyond traditional source-level testing and verification practices.

\begin{table}[h]
\caption{Illustrative Quotes by Type of Issue}
\label{tab:issue-types}
\centering
\begin{tabular}{p{8cm}}
\toprule
\textbf{Code Level} \\
\midrule
\textit{Logical: "Instead of the prompt returning a message, it would instead crash the entire logic due to an error being returned (dupe key error) in our backend logic."} (P040) \\\\

\textit{Interface: "Main issues identified during testing were how the LLM was outputing its response, how it generated buy links and how it was pricing certain PC components."} (P005) \\\\

\textit{Syntax: "However, during testing, we noticed that the LLM would sometimes not give back the response in the format we wanted, and the description would be too small/vague sometimes."} (P037) \\\\

\textit{Teamwork: "One main issue identfied during testing was sometimes the bugs that were in the responses created by the LLM was hard for another person to re-create easily."} (P074) \\\\

\midrule
\textbf{Model Behaviour} \\
\midrule
\textit{LLM Output: "The LLM would end up hullucinating, providing ingredients that were not in the picture, or did not provide recipes strictly related to the ingredients provided."} (P013) \\\\

\textit{Prompt Sensitivity: "Responses began as very stagnant as the LLM would suggest the same outfit every single time no matter what the prompt was."} (P001) \\
\bottomrule
\end{tabular}
\end{table}

\subsection{Challenges in Testing LLM-Powered Systems}

In testing LLM-powered systems, teams encountered challenges that varied in their relationship to source-level testing and verification. Some issues were tightly coupled with the system’s code and required backend debugging, structured test logic, or integration validation. Others were only partially related, involving prompt construction and formatting tasks implemented in code. A third category involved challenges that arose primarily from the LLM’s behavior and had limited connection to the source code. What follows is a description of the challenges most frequently reported, grounded in the participants' own accounts, organized by their relevance to source-level tests. Table \ref{tab:challenges} presents illustrative quotes as evidence of the findings.

The most directly connected challenge was \textit{integration issues}, which emerged when teams attempted to connect LLM outputs with backend logic, databases, or frontend components. Participants described problems such as mismatches in data format, API failures, and unhandled exceptions when the model’s output deviated from expected structures. Debugging these issues required teams to write or revise code to validate JSON structures, parse model responses, or handle asynchronous requests. Integration bugs often surfaced during deployment or full-system testing, demanding adjustments to both logic and error-handling routines.

Also related to source-level testing was \textit{non-deterministic behavior}, which complicated the use of traditional test cases. Although the core issue was the model’s stochastic nature, it had a direct impact on the structure of source-level tests. Teams described writing validation logic that allowed for variability in content while enforcing correctness of structure, and some implemented retry mechanisms or fallback behaviors when responses failed to meet format expectations. This required modifications in test design and backend implementation to accommodate unpredictable outputs.

A third challenge, partially tied to source-level work, involved \textit{prompt engineering difficulties}. Teams discovered that small changes to prompts could cause significant shifts in model behavior, and some of these prompts were generated or modified dynamically within the code. Debugging prompt construction involved testing prompt templates, managing conditionals, and validating input processing logic to ensure consistent LLM behavior across scenarios.

Less directly connected to source-level testing was \textit{hallucinations}. Teams frequently reported that the model produced factually wrong, fabricated, or incoherent content, even when prompts were precise. These errors often could not be traced back to a specific code issue and required teams to rely on manual exploratory testing, user reviews, or prompt refinement to detect and manage output problems. Automated testing methods were rarely applicable to this type of issue.

\textit{Output imbalance and the risk of biased outcomes} was another challenge grounded in content behavior rather than code. Teams noted that the model consistently favored certain patterns, such as always selecting the same correct answer in multiple-choice quizzes, over-relying on popular items in recommendations, or mishandling edge cases in user inputs. While not overtly harmful, these biases led to concerns about misleading results. Responses to this issue typically focused on adjusting prompts or adding content-level constraints rather than modifying source code.

These varied challenges required teams to rethink conventional testing assumptions. Rather than relying solely on fixed outputs or predefined test scenarios, teams had to adopt flexible evaluation strategies that could accommodate uncertainty, partial correctness, and evolving behaviors. This shift pushed testing beyond a purely technical exercise and into a space where verification also involved human judgment, contextual reasoning, and continuous refinement.

\begin{table}[h]
\caption{Illustrative Quotes by Challenge}
\label{tab:challenges}
\centering
\begin{tabular}{p{8cm}}
\toprule
\textbf{Integration Issues} \\
\midrule
\textit{"our backend were not returning data to our frontend correctly (…) we had to develop new testing strategies (manual prompt evaluation) to access the LLM data before it reached the frontend…"} (P007) \\ \\

\textit{"(…) came down to how the chatgpt and gemini's api key specifically was loaded in Jest which took several hours to fix."} (P013) \\

\midrule
\textbf{Non-Deterministic Behavior} \\
\midrule
\textit{"Conventional testing techniques like unit tests were not effective for LLM features due to their non-deterministic nature. Even for the same input there could be different outputs…"} (P002) \\ \\

\textit{"LLM responses aren't always the same. That unpredictability made automated testing a bit unreliable…"} (P003) \\

\midrule
\textbf{Prompt Engineering Difficulties} \\
\midrule
\textit{"…took us a while to figure out the right set of prompts to ask the LLM that would deliver the best result."} (P011) \\ \\

\textit{"none of us really used LLMS consistently so we had to understand exactly what the AI looks for when it gets the prompt and how we can manipulate the prompt to only give the appropriate answer related to our stuff."} (P090) \\

\midrule
\textbf{Hallucinations} \\
\midrule
\textit{"…math is currently something that will provide hallucinations as answers, so further propmt ir logic enginering will be needed."} (P024) \\ \\

\textit{"The challenges in the LLM is very diffcult beacuse there is so many time that the AI cannot answer the question or may not know… most of the time it would just give you something that is not even the same topic as the question."} (P070) \\

\midrule
\textbf{Output Imbalance and Bias} \\
\midrule
\textit{"…there is always the bias risk that the model pulls information from the internet which is entirely inaccurate."} (P085) \\ \\

\textit{"espically when it is a lower level model, all testers required to be patient and determined to solve the issues as many times the LLMs revert back to their problematic state thus the testers needed extra testing to ensure alfter the solutions were implemented"} (P079) \\
\bottomrule
\end{tabular}
\end{table}

\section{Discussion}
\label{sec:discussion}
This section discusses the key findings of our analysis by examining how participants approached the testing of LLM-powered systems. We begin by answering the central research question of the study, summarizing the testing strategies and challenges reported in the participant projects. After establishing this foundation, we explore three core aspects of the testing process in greater depth, highlighting how students combined source-level reasoning with adaptive practices to manage the unpredictability of LLM outputs. We then relate these findings to existing research, identifying points of convergence and extension. The section concludes with reflections on the implications of our results for both the research community and practitioners building LLM-integrated applications. It is important to interpret the discussion in the context of the scope of the study: The findings are based on student-led development in a university course. While the projects simulated realistic development constraints, the insights reflect how emerging practitioners engage with LLM testing and may not fully represent practices in professional or industrial environments.

\rqone[
    \tcblower
    \textbf{Answer:} LLM-powered systems are tested using a combination of traditional and adaptive strategies that span both source-level and system-level concerns. At the source level, teams engage in backend debugging, prompt logic refinement, and integration fixes to manage issues such as format mismatches, API failures, and incorrect response handling. Unit and integration tests are used to validate backend functions and ensure data flow between components. At the system level, teams rely on exploratory, scripted, and ad hoc testing to assess model behavior, especially in the face of non-deterministic outputs, hallucinations, and prompt sensitivity. Challenges include the variability of model responses and the difficulty in determining whether outputs are incorrect or simply misaligned with expectations. This ambiguity led teams to rely on judgment-based evaluations rather than predefined outcomes. Overall, testing LLM-powered systems demands a layered approach where source-level reasoning supports system-level adaptation, enabling teams to manage the unpredictability and complexity introduced by generative components.
]{}

\subsection{Navigating a New Development and Testing Landscape}

Coding and testing LLM-powered systems introduced teams to an environment that extended beyond conventional practices. While traditional techniques such as unit testing, integration testing, and scripted verification remained relevant, they proved insufficient to fully capture the complexities introduced by LLM integration. Teams had to deal with variable model outputs, difficulties in defining correct behavior, and limited means to reproduce issues that were not tied to deterministic code defects.

To address these scenarios, the teams adopted adaptive strategies that blended deterministic coding practices with exploratory and behavior-driven testing. Verifying LLM behavior was not solely a technical task but also a judgment-based process that required context-aware reasoning, flexible criteria for success, and tolerance for ambiguity. Below, we expand on how these practices emerged and evolved across the three core areas identified.

\textbf{Testing Grounded in Source Code Analysis.} Despite the presence of a non-deterministic model component, the foundation of LLM-powered system development remained rooted in source code analysis. Teams engaged in backend debugging to trace issues with prompt preparation, API integration, and response handling. They reviewed requests and adjusted parameters to diagnose failures, often working through back-end modules responsible for analyzing and structuring LLM inputs and outputs.

Even when model behavior appeared problematic, teams often began by validating the underlying infrastructure to rule out logic, interface, or syntax errors. This revealed practices grounded in source-level reasoning, even when part of the system lies beyond their control. As such, debugging LLM-powered systems requires a detailed understanding of how code and model behavior interact, making traditional software engineering skills essential to navigate these systems.

\textbf{Hybrid Strategies as the Norm in Coding and Testing.} Rather than relying on a single testing strategy, most teams adopted hybrid approaches that combined exploratory testing, scripted test cases, and selective automation. Exploratory testing was particularly common, used to probe how the LLM reacted to varying inputs and edge cases. Scripted testing helped verify known behaviors, while unit and integration tests supported systematic validation of code infrastructure.

This strategic layering reflected the dual nature of LLM-based applications. On one hand, students needed to ensure that backend functions worked as intended. On the other hand, they had to explore how generative outputs aligned with expectations, often without a clear oracle to guide them. The result was a testing process that spanned structured and improvised techniques.

\textbf{Coping with Unpredictability in LLM Behavior.} The unpredictability of LLM responses emerged as a persistent challenge during both coding and testing. Teams encountered issues such as hallucinations, unstable outputs, and prompt sensitivity, none of which could be reliably detected or resolved using conventional assertions or test cases. These issues could not be resolved through traditional debugging. Testing under these conditions required flexible evaluation criteria, fallback logic, and tolerance for partial correctness.

To manage this uncertainty, teams adjusted their validation strategies. Some implemented fallback logic to handle invalid responses, while others built tests that prioritized structure and format over exact content. In several cases, testing relied on qualitative evaluations rather than binary pass/fail criteria. This shift highlights a key characteristic of LLM testing: correctness becomes probabilistic, and system validation must account for variance in model output even when the surrounding infrastructure is stable.

\subsection{Comparing Findings with the Literature}
\label{sec:discussion-comparison}

Recent studies on LLM-powered systems have largely focused on characterizing model behavior, including its accuracy, robustness, and limitations in various domains~\cite{lin2024arxiv, feng2024llmeffichecker, yang2024evaluation, bedi2024systematic}. While these contributions provide a foundation for understanding LLM performance, they tend to focus on output quality rather than the verification and validation practices adopted during system development. In contrast, our study focuses on how LLM-powered applications are tested as part of broader software systems, including the interaction between prompts, code, and backend logic.

Across our collected reports, participants relied on exploratory testing, unit tests, and scripted validations to assess system behavior. Similar hybrid testing patterns have been observed in broader LLM development efforts, where teams employ a combination of human-driven interaction and structured prompts to evaluate model responses~\cite{chen2025design}. However, existing studies tend to highlight these strategies in isolated evaluations or as part of prompt iteration, whereas our participants integrated them directly into the testing cycle of full applications.

Non-determinism and prompt sensitivity were frequently reported by participants as challenges that affected test design. These issues have also been identified in controlled experiments that measure how model outputs vary across different prompt styles and API versions~\cite{ma2024my, chen2025empirical}. While prior work describes prompt instability as a limitation of LLM usage, our data shows how it influences code-level decisions: participants implemented retries, adjusted parsing logic, and wrote fallback routines to account for variability. These behaviors suggest that prompt stability is not only a usability concern but also a testing constraint embedded in system logic.

Issues related to formatting inconsistencies, interface errors, and model integration failures were also common in our study. Similar problems are mentioned in papers focused on LLM deployment in real-world environments, particularly when models are embedded in multi-component systems~\cite{hou2024large, morales2024dsl, zheng2025towards}. However, much of that literature describes system-level integration abstractly, whereas our participants identified specific parsing bugs and logic breakdowns that occurred when models returned outputs that did not match expected formats. These reports support recent calls for more robust validation strategies and structured testing layers when working with LLM APIs~\cite{zapkus2024unit, asgari2024testing}.

Some participants described difficulty in deciding whether an LLM output was incorrect or merely unexpected. This aligns with concerns in the literature about the ambiguity of evaluation in generative systems~\cite{chen2025design}. While prior work points out the absence of fixed oracles in model evaluation, our findings show how that uncertainty appears in practice: testers debated acceptable variations, adjusted their test logic accordingly, and often deferred judgment to manual review.

Although several studies propose new frameworks for testing LLMs—such as prompt coverage metrics or adversarial probing~\cite{feng2024llmeffichecker, yang2024evaluation}—they typically assess models in isolation. In our study, testing was grounded in the day-to-day challenges of making a system work end-to-end. Participants worked through integration failures, and inconsistent behavior, often with limited tools to anticipate or control model output. This highlights a different perspective: testing was not only about evaluating what the model could produce, but about ensuring that its responses were usable, parseable, and compatible with the broader system.

In summary, our findings reflect several concerns raised in the literature, including non-determinism, prompt instability, and evaluation uncertainty. However, they also show how these concerns translate into concrete verification behaviors within the software development lifecycle. Rather than focusing solely on model performance or prompt design, this study contributes a view of testing that is shaped by development constraints, integration tasks, and adaptive strategies embedded in source-level code.

\subsection{Implications to Research}
\label{sec:implications-research}

This study contributes to research on how LLMs are developed and how systems that use LLMs are implemented. Our findings highlight that LLM integration is not limited to connecting an API or designing prompts: it involves adapting software engineering practices throughout the development cycle. In this sense, we offer an overview of how software teams might navigate prompt instability, handle output unpredictability, and modify backend logic to manage LLM behavior.

Our research indicates that successful implementation hinges on treating the LLM as an integral component within the system architecture, rather than merely a passive service. Developers debug prompt logic in code, adjust system design to support fallback behavior, and use layered testing approaches to evaluate both internal functions and external responses. These practices are currently underexplored in empirical software engineering research; however, they are crucial for real-world LLM development.

This work also opens up space for new research directions that focus on the system, not just the model. Future studies can build on these results to investigate how development teams reason about correctness in probabilistic environments, how they structure testing workflows around prompt and response expectations, and how quality assurance evolves when outputs are only partially predictable.

In addition, since our study utilizes student-authored artifacts, this approach can be extended to other domains or professional teams, enabling researchers to track decision-making over time and compare different strategies for managing the interaction between generative models and software systems.

\subsection{Implications to Practice}
\label{sec:implications-practice}

This study offers practical guidance for software teams working with LLM-powered systems. First, it emphasizes that testing LLMs is not only about checking output accuracy; it also involves validating how model responses are handled and displayed by the rest of the system. Many teams encountered errors not because the model failed but because the system could not properly process its output. We recommend that developers include validation steps that check for format, type, and completeness before integrating model responses into other components.

Second, prompt engineering should be treated as part of software development, not as a one-time design activity. Prompts often require iteration, debugging, and refinement during development and maintenance. Teams can benefit from assigning responsibility for prompt logic, documenting prompt structures, and testing prompts under different conditions. We believe that tools that support prompt versioning, testing in isolation, and behavioral monitoring would be especially valuable in this context.

Third, testing workflows should accommodate non-determinism. Traditional test cases may not be sufficient for LLM responses, especially when variability is expected or even desired. We suggest that developers consider writing partial or structure-based assertions, using retry mechanisms, and planning for exploratory testing when outputs can vary. Teams might also benefit from defining tolerance thresholds or response expectations that account for acceptable variation rather than strict correctness.

Finally, our study reveals that ambiguity is inherent in LLM testing. In many cases, teams were unsure whether outputs were truly incorrect or simply diverged from expectations. Accepting this uncertainty as part of the testing process—and creating space for manual review or collaborative evaluation—can help teams avoid over-testing or misclassifying useful behavior as failure. We suggest that teams designate specific roles or checkpoints for reviewing model outputs during development. Although these insights stem from student projects, they illustrate emerging challenges in LLM testing and highlight areas where both tooling and educational support may be needed. We recommend that future research continue to explore how such practices scale or differ in professional development settings.

\section{Threats to Validity}
\label{sec:limitations}

A key threat to this study lies in the use of students working in a simulated real-world setting. While students are not industry professionals, the course was designed to reflect realistic development constraints. Teams were required to deploy functional applications with a SQL database, a publicly accessible endpoint, and a working integration with an LLM (e.g., ChatGPT, Gemini). Testing activities were explicitly part of the instruction and evaluation criteria. Students made independent testing decisions and documented their experiences through individual reports. These conditions enabled us to observe testing practices in a context that, while educational, captures several features of real-world system development.

Second, while the projects were conducted in a realistic academic setting, the participants were students rather than professional developers. Their testing practices may reflect early-stage reasoning rather than established industrial procedures. However, we believe this offers a valuable lens into how future practitioners are learning to work with LLMs under realistic constraints.

Another limitation is the question of generalizability. As a qualitative case study, we do not claim statistical generalization. Instead, our goal is transferability. We provide detailed descriptions of the setting, tasks, and participant reflections, enabling readers to assess whether our findings may be applicable in other contexts. All 99 reports were analyzed in full, without sampling, and quotations are included throughout the findings to support transparency and credibility. This aligns with established qualitative standards, where depth, richness, and clarity of context are prioritized over broad generalization.

We also recognize the risk of researcher bias during data analysis. To reduce this threat, we used a multi-step thematic analysis process. A structured, prompt-guided, model-assisted coding approach was used, and two researchers independently verified and refined all extracted data. Only quotations explicitly written by participants were included; no inferences were made. All disagreements were resolved through discussion. This approach helped ensure consistency, traceability, and rigor in the interpretation of the data.

We selected ChatGPT 4.0 for initial code suggestion to accelerate the open coding process and reduce redundant labeling. Prompts were carefully calibrated by the first author using test excerpts to ensure reliable topic extraction and consistent terminology. All generated codes were reviewed by two researchers independently and validated against the original context. Disagreements were resolved through discussion. We did not rely on the LLM to generate themes or interpret data. Instead, ChatGPT served as a tool to support focused coding, with final coding and thematic synthesis performed manually by the research team. While this mitigated the risk of bias or circular reasoning, future studies should continue to evaluate the role of LLMs in supporting or shaping qualitative research practices.

Finally, the context-specific nature of this study, which focuses on a single course at a single institution, may limit its scope of applicability. However, this is consistent with the goals of exploratory case studies, which aim to understand emerging or underexplored phenomena in depth. Our study focuses on how software teams adapt testing practices when integrating LLMs, offering insights that can inform future research and practice, even if findings are not universally generalizable.

\section{Conclusions} 
\label{sec:conclusions}

This study explored how students test LLM-powered software systems in a realistic development context. By analyzing 99 individual reports from deployed projects, we uncovered how testing extended beyond conventional verification techniques to accommodate the non-deterministic and often unpredictable behavior of large language models. Despite working in a classroom setting, participants engaged in structured testing efforts, combined manual and automated strategies, and adjusted system logic to handle variable outputs.

These efforts were grounded in source-level testing practices. Even though exploratory and ad hoc testing formed the core of the testing process, they were frequently used in combination with unit testing to validate backend functions and support system integration. In several cases, exploratory testing was used to observe prompt behavior, while unit tests helped verify the stability of processing logic. Students adjusted the backend code to handle LLM output variability by refining the parsing logic, adding checks for expected formats, and implementing fallback behaviors. These adjustments show that prompt handling was not limited to manual inspection but was actively managed within the code through structured testing and defensive programming practices.

Therefore, our findings connect to core concerns in source code analysis and manipulation, such as logical error detection, interface coordination, and runtime adaptation. Many of the issues reported by students (e.g., broken backend logic, incorrect handling of LLM responses, and formatting problems) were only discovered by checking and changing the code directly. At the same time, challenges like hallucinations and prompt sensitivity forced teams to move beyond fixed test cases and build systems that could deal with unpredictable outputs. These results demonstrate how LLM-powered systems are tested in practice, combining source-level reasoning with adaptive strategies to manage the complexity of generative behavior.

\ifCLASSOPTIONcaptionsoff
  \newpage
\fi

\balance
\bibliographystyle{IEEEtran}
\bibliography{bib.bib}

\end{document}